# High critical-current density with less anisotropy in BaFe$_2$(As,P)$_2$ epitaxial thin films: Effect of intentionally grown *c*-axis vortex-pinning centers

Hikaru Sato[1], Hidenori Hiramatsu[1, 2], Toshio Kamiya[1, 2], and Hideo Hosono[1, 2, 3 *]

1: *Materials and Structures Laboratory, Tokyo Institute of Technology, Mailbox R3-1, 4259 Nagatsuta-cho, Midori-ku, Yokohama 226-8503, Japan*

2: *Materials Research Center for Element Strategy, Tokyo Institute of Technology, Mailbox S2-16, 4259 Nagatsuta-cho, Midori-ku, Yokohama 226-8503, Japan*

3: *Frontier Research Center, Tokyo Institute of Technology, Mailbox S2-13, 4259 Nagatsuta-cho, Midori-ku, Yokohama 226-8503, Japan*

Abstract

We report herein a high and isotropic critical-current density $J_c$ for $BaFe_2(As,P)_2$ epitaxial films. The isotropy of $J_c$ with respect to the magnetic-field direction was improved significantly by decreasing the film growth rate to 2.2 Å/s. The low growth rate served to preferentially align dislocations along the *c*-axis, which work well as *c*-axis vortex-pinning centers. Because of the intentional introduction of effective pinning, the absolute $J_c$ at 9 T was larger than that obtained for other iron-based superconductors and conventional alloy superconducting wires.

Footnote:

*: Author to whom correspondence should be addressed. Electronic mail: hosono@msl.titech.ac.jp

Iron-based superconductors,[1] with their high critical temperature ($T_c$) up to 55 K[2] and high upper critical field ($H_{c2}$) up to 100 T,[3] are seen as new candidate materials for high-magnetic-field applications such as superconducting wires and tapes. In particular, 122-type Ba(Fe$_{1-x}$Co$_x$)$_2$As$_2$ (BaFe$_2$As$_2$:Co) has been found to have attractive properties such as a large $H_{c2}$ of over 50 T, a small anisotropy factor ($\gamma$ = 1 to 2),[4] and an advantageous grain-boundary nature.[5] Thus, of the iron-based superconductor thin films, BaFe$_2$As$_2$:Co epitaxial films have been investigated most intensively, and their self-field critical-current density $J_c$ is currently far above 1 MA/cm$^2$.[6]

However, as compared to BaFe$_2$As$_2$:Co, 122-type BaFe$_2$(As$_{1-x}$P$_x$)$_2$ (BaFe$_2$As$_2$:P) exhibits a higher $T_c$ (~31 K) and a comparable $\gamma$.[7] In addition, it was recently reported[8] that BaFe$_2$As$_2$:P epitaxial films have a very high $J_c$ of 10 MA/cm$^2$, which is two to three times greater than that of BaFe$_2$As$_2$:Co.[9,10] Unfortunately, the decay rate of $J_c$ for a BaFe$_2$As$_2$:P epitaxial film under a magnetic field[7] is greater than that of a BaFe$_2$As$_2$:Co epitaxial film.[11] This observation is attributed to the weak vortex pinning in BaFe$_2$As$_2$:P because an isovalent P dopant is not thought to work as an effective pinning center.[12,13] This characteristic contrasts sharply with BaFe$_2$As$_2$:Co, in which the disorder of aliovalent Co dopants in the Fe layers works as intrinsic vortex pinning centers. Therefore, to exploit the advantages of BaFe$_2$As$_2$:P films, it is vital to enhance their vortex pinning so that a high $J_c$ can be maintained under high magnetic fields.

Another important issue for wire/tape applications is the anisotropy of $J_c$. Since high-$T_c$ superconductors such as cuprates and 1111-type iron-based superconductors have distinct layered structures, $H_{c2}$ with $H$ // $ab$ is higher than $H_{c2}$ with $H$ // $c$ because of their electronic anisotropy; their $J_c$ properties also reflect this intrinsic crystallographic anisotropy. Therefore, introducing pinning centers along the *c*-axis is a practical way to reduce their anisotropy.[14,15] Heavy-ion irradiation experiments have clarified the attractive properties of iron-based superconductors, which are totally different from cuprates; introducing a high density of nanosize columnar defects does not degrade $T_c$ until the dose-matching field becomes 21 T, which indicates the excellent potential of iron-based superconductors for high-field applications.[16] Although these results revealed the high potential of iron-based superconductors if pinning centers as structural defects are introduced, it is difficult to apply to large-scale superconducting wires and tapes. Therefore, a simple fabrication process, which controls the shape, size, and density of defects, is more practical than introducing artificial pinning centers.

In the present study, we obtained a large isotropic $J_c$ in $BaFe_2As_2$:P epitaxial films grown by high-temperature pulsed-laser deposition (PLD) at 1050 °C. The results indicated that the characteristics of vortex pinning depend largely on growth rate, which affects the structure of microscopic defects in the epitaxial films.

$BaFe_2As_2$:P films of 150–200 nm thickness were grown on MgO (001) single-crystal substrates by PLD. Optimally P-doped polycrystalline $BaFe_2(As_{0.7}P_{0.3})_2$ disks were used as the PLD targets, which were synthesized by a solid-state reaction of BaAs + $0.4Fe_2As$ + $0.6Fe_2P$ → $BaFe_2(As_{0.7}P_{0.3})_2$ at 930 °C for 16 h in a sealed Ar-filled stainless-steel tube. The pulsed-laser excitation source was the second harmonic of an Nd:YAG laser (wavelength = 532 nm, repetition rate = 10 Hz, laser fluence ~ 3 J/cm$^2$). The base pressure of the growth chamber was approximately $5 \times 10^{-7}$ Pa. Instead of a halogen-lamp heater,[17] a semiconductor infrared laser diode was used (wavelength =

975 nm, maximum power = 300 W) to achieve high substrate temperatures ($T_s$) up to 1400 °C. The $T_s$ was calibrated by an *ex-situ* measurement using a thermocouple connected to a substrate directly. To heat the substrate, a Mo plate was tightly contacted to the backside of the substrate and heated by focused infrared light, which provided uniform $T_s$ over the substrate. First, we varied $T_s$ from 700 to 1300 °C by maintaining similar growth rates of ~3 Å/s, which is almost the same as that used for the epitaxial growth of $BaFe_2As_2$:Co.[17] After finding the optimum $T_s$, the growth rate was varied from 2.1 to 3.9 Å/s by changing the substrate–target distance.

To determine the crystalline phases and the small amount of impurity phases, $\theta$-coupled $2\theta$-scan x-ray diffraction measurements (XRD, anode radiation: Cu K$\alpha$) were performed with a high-power conventional XRD apparatus. The crystallinity of the epitaxial films was characterized on the basis of full widths at half maximum (FWHM) of the out-of-plane ($2\theta$-fixed $\omega$ scans) rocking curves ($\Delta\omega$) of 002 diffraction and the in-plane ($2\theta_\chi$-fixed $\phi$ scans) rocking curves ($\Delta\phi$) of 200 diffraction with a high-resolution XRD apparatus [HR-XRD, CuK$\alpha_1$ monochromated by Ge (220)]. The chemical composition was determined from electron-probe microanalyzer (EPMA). For quantitative analyses, we employed the atomic number, absorption, fluorescence (ZAF) correction method using the following standard samples; $BaTiO_3$ for Ba, Fe for Fe, LaAs for As, and InP for P, which provided reasonable results as will be shown later (for a $BaFe_2As_2$:P film grown at an optimum $T_s$ ~ 1050 °C, Ba: Fe: (As + P) = 1: 2.1 : 2.0).

The surface morphology was observed with an atomic force microscope (AFM). Cross-sectional microstructural images were obtained by scanning transmission electron microscopy (STEM). The variation in chemical composition around defects was measured using energy-dispersive x-ray spectroscopy (EDXS) with a spatial resolution of about 1 nm. These characterization were performed at room temperature.

Temperature ($T$) dependence of electrical resistivity ($\rho$) was measured by the four-probe method using Au electrodes with a physical property measurement system. To examine the transport properties of $J_c$, the films were patterned into microbridges by photolithography and an Ar milling process. The transport $J_c$ was determined from voltage–current curves with the criterion of 1 μV/cm under an external magnetic field ($H$) of up to 9 T. The angle $\theta_H$ of the applied $H$ was varied from −30 to 120° (0 and 90° corresponding to $H$ // $c$-axis normal to the film surface and $H$ // $ab$ plane, respectively).

Figure 1(a) summarizes the relationship between the growth condition ($T_s$ and laser fluence) and crystallographic orientation of BaFe$_2$As$_2$:P thin films grown at growth rates of ~3 Å/s. When $T_s \leq 950$ °C ["Random oriented" region: triangles in Fig. 1(a) and lower XRD pattern in Fig. 1(b)], the films were preferentially oriented along the $c$-axis along with a smaller portion of nonoriented BaFe$_2$As$_2$:P crystallites and Fe impurities appeared. When $T_s$ was increased to 1100 °C ["Epitaxial" region: circles in Fig. 1(a)], epitaxial BaFe$_2$As$_2$:P films were obtained; that is, only 00$l$ diffractions were observed out of plane [middle row of Fig. 1(b)], and due to the tetragonal lattice, a clear four-fold symmetry was observed in the in-plane $\phi$ scan [Fig. 1(c)]. At $T_s \geq 1100$ °C ["00$l$ & $hh$0" region, squares in Fig. 1(a) and the top XRD pattern in Fig. 1(b)], 00$l$ and $hh$0 preferential orientations were observed, which is similar to that observed for high-$T_s$ growth of BaFe$_2$As$_2$:Co.[17] A weak Fe 002 peak is observed in the XRD patterns. We confirmed that impurity Fe particles segregated in the bulk regions of the films, not at the film–substrate interface (see Supplementary Figures S1(a) and (b) for the cross-sectional bright-field STEM image and EDXS spectra)[18], sharply different from the result of BaFe$_2$As$_2$:Co films grown by PLD using a KrF excimer laser[19]. In the latter report, the Fe impurity is epitaxially grown at the interface. However, similar segregation in the bulk region is observed also in BaFe$_2$As$_2$:Co epitaxial films fabricated using a Nd:YAG laser as an excitation source [S1(c) and (d)][18].

Based on these results, we concluded that the optimum $T_s$ for epitaxial growth of BaFe$_2$As$_2$:P is ~1050 °C. The chemical composition of the film, as determined by EPMA, was Ba: Fe: As: P = 19.7: 41.5: 30.3: 8.5 [i.e., P/(As + P) = 22%], indicating that the P doping level was lower than the nominal composition of the PLD target [P/(As + P) = 30%]. The lattice parameters of the film were $a = 3.95$ Å and $c = 12.830$ Å, which differ slightly from those of a single crystal with the same chemical composition ($a = 3.94$ Å and $c = 12.89$ Å)[20]. This result implies that a tensile strain in the *ab* plane is introduced into the epitaxial film because of the larger in-plane lattice parameter of the MgO substrate ($a = 4.21$ Å).

The surface morphology of the film grown at the optimum $T_s$, as observed by AFM, shows that the film grows via spiral-island growth and has a smooth surface with the step height corresponding to half of the *c*-axis length [Fig. 1(d)]. Note that there are few droplets and pit structures, although these are often observed in BaFe$_2$As$_2$:Co epitaxial films.[21] These results would originate from the higher $T_s$ growth (1050 °C) than that of BaFe$_2$As$_2$:Co epitaxial films (800–850 °C),[17] which promotes the migration of deposition precursors and reconstruction of their structure at the growing surface. Such an atomically flat surface would be useful for future applications such as multilayer junction devices.

Figure 1(e) shows $\rho$ as a function of $T$ for a BaFe$_2$As$_2$:P epitaxial film grown at the optimum $T_s$ = 1050 °C. A sharp superconducting transition at $T_c$ = 26.5 K was observed with a small transition width of $\Delta T_c$ = 1.5 K, which is consistent with the slightly underdoped P-concentration determined by EPMA [P/(As + P) = 22%]; however, this $T_c$ is 15 K higher than that of a single crystal with the same P concentration and is comparable to that of a 27% P-doped single crystal.[22] As mentioned above, a tensile strain in the *ab* plane is introduced into the BaFe$_2$As$_2$:P epitaxial films, which suggests that the effect of the tensile strain is opposite to an optimally doped BaFe$_2$As$_2$:Co epitaxial film because a tensile strain decreases $T_c$ for BaFe$_2$As$_2$:Co[23]. On the other hand, it was recently reported that a tensile strain in a BaFe$_2$As$_2$:P epitaxial film shifts its optimum doping level to an underdoped region[24]. Thus, this result would be understood that the tensile strain shifts the optimum doping level to an underdoped region and the $T_c$ of the present underdoped (P/(As + P) = 22%) BaFe$_2$As$_2$:P epitaxial films were enhanced.

After optimizing the $T_s$, the growth rate was varied with the $T_s$ held at the optimum $T_s$ = 1050 °C. As seen in Fig. 1(f), the rocking-curve FWHMs of $\Delta\omega$ and $\Delta\phi$ remain almost constant at ~ 0.5° for a growth rate ranging from 2.1 to 3.9 Å/s. This trend is different from that for BaFe$_2$As$_2$:Co epitaxial films,[25] where low growth rates such as 2.1 Å/s led to large $\Delta\omega$ and $\Delta\phi$ (FWHM ~ 1°). This finding means that BaFe$_2$As$_2$:P films may be grown over a wider range of growth rates, a result that is also attributed to the higher optimum $T_s$.

Next, we discuss anisotropy in transport $J_c$. For several growth rates, Fig. 2 shows angular dependence of $J_c$ for BaFe$_2$As$_2$:P epitaxial films grown at the optimum $T_s$. The data show that $J_c$ increases with decreasing growth rate over the entire range of $\theta_H$. This result is attributed to the improvement in the self-field $J_c$; namely, at 12 K, the self-field $J_c$ increased from 2.70 to 5.14 MA/cm$^2$ as the growth rate decreased from 3.9 to 2.2 Å/s. All the films exhibit intrinsic $J_c$ peaks at $\theta_H = 90°$ ($H \,//\, ab$), which is in agreement with the results for the film grown at the higher growth rate of 5.0 Å/s.[7] The present films, however, exhibit additional $J_c$ peaks at $\theta_H = 0°$ ($H \,//\, c$). The value of $J_c$ at $\theta_H = 0°$ ($J_c^{H//c}$) increases with decreasing growth rate, then it exceeds $J_c$ at $\theta_H = 90°$ ($J_c^{H//ab}$) at the growth rate of 2.2 Å/s. These results indicate that vortex-pinning centers along the $c$-axis are introduced into BaFe$_2$As$_2$:P epitaxial films when these films are grown at a lower growth rate. Note that the $J_c$ peak at $\theta_H = 0°$ becomes larger and sharper upon decreasing the growth rate, which averages out the angle dependence of $J_c$ and contributes to the highly isotropic property. These results indicate that the vortex-pinning properties and anisotropy of BaFe$_2$As$_2$:P epitaxial films can be controlled by tuning the growth rate.

Figure 3(a) compares the dependence of $J_c^{H//ab}$ (closed symbols) and $J_c^{H//c}$ (open symbols) at 4 K for BaFe$_2$As$_2$:P epitaxial film grown at the optimum $T_s$ (1050 °C) and growth rate (2.2 Å/s) with what is found for other iron-based superconductor epitaxial films exhibiting high $J_c$. The self-field $J_c$ of the optimally grown BaFe$_2$As$_2$:P epitaxial film was over 7 MA/cm$^2$, which is comparable to the best results recently reported,[8] and a high $J_c^{H//ab}$ of over 1 MA/cm$^2$ was maintained even at 9 T. It would be noteworthy the value of $J_c^{H//ab}$ obtained at 9 T = 1.1 MA/cm$^2$ (corresponding to a pinning force of 99 GN/m$^3$) is the highest obtained for iron-based superconductor thin films, as shown in Fig. 3(a).[26–32] Moreover, at 9 T, $J_c^{H//c}$ is 0.8 MA/cm$^2$ and the pinning force is 72 GN/m$^3$.

These values are higher than the previously reported ones, including those for 122-type compounds [e.g., $BaFe_2As_2$:P films with artificial pinning centers created by $BaZrO_3$ nanoparticles[26] and oxygen-rich $BaFe_2As_2$/$BaFe_2As_2$:Co superlattice (SL) films[28]] and conventional alloy superconducting wires (Nb-Ti,[33] $Nb_3Sn$,[34] and $MgB_2$[35]).

Figure 3(b) compares the $J_c$ anisotropy of the $BaFe_2As_2$:P epitaxial film with the $J_c$ of other 122-type films[7,26,28,29] in similar ranges of $H$ and $T$. The $BaFe_2As_2$:P epitaxial film obtained in this study exhibited a lower anisotropy than those of $BaFe_2As_2$:P[7] and $BaFe_2As_2$:Co/Fe buffer.[29] Moreover, the anisotropy obtained is much lower than that of the SL thin films,[28] and is comparable to that of the $BaFe_2As_2$:P films with artificial pinning centers created by $BaZrO_3$ nanoparticles.[26]

As discussed above, the strong vortex pinning and the isotropic $J_c$ properties were achieved by decreasing the growth rate for the $BaFe_2As_2$:P epitaxial films. Therefore, we investigate the origin of the pinning centers by microstructure analysis. Figures 4(a) and 4(b) show cross-sectional bright-field STEM images of $BaFe_2As_2$:P epitaxial films grown at 2.2 and 3.9 Å/s, respectively. Although small island structures ($\sim$ 20 nm in lateral size) rich in oxygen are periodically found at the heterointerface in the film grown at 2.2 Å/s, as indicated by the slanted black arrows, sharp heterointerfaces without a reaction layer were observed throughout the region. Other planar or line defects in the *ab* plane, such as stacking faults, were not detected. However, as indicated by the vertical white arrows in Fig. 4, many vertical defects, which occur at a higher density compared with that found for $BaFe_2As_2$:Co films,[36] are observed in the STEM images. The number of defects did not significantly differ between the $BaFe_2As_2$:P films grown at 2.2 Å/s and 3.9 Å/s, but the shape and microstructure of the defects did differ. Most of the defects in the film grown at 2.2 Å/s start appearing at midthickness and are oriented parallel to the *c*-axis, which would be assigned to vertical dislocations. However, most of the defects in the film grown at 3.9 Å/s originate at the heterointerface just at the substrate surface and are tilted with respect to the *c*-axis. Such defects could be induced by lateral growth of the epitaxial domains. These results suggest that the defects in the film grown at 3.9 Å/s are domain boundaries that initiate at the onset of film growth. This difference in the structure of the defects would correspond to the results of Fig. 2: against a magnetic field with $\theta_H = 0°$, the vertical dislocations would pin vortices more strongly than the tilted domain boundaries, resulting in the sharper peak near $\theta_H = 0°$ for films grown at 2.2 Å/s and the broad peak near $\theta_H = 0°$ for films grown at 3.9 Å/s.

The STEM-EDXS line scans [Figs. 4(c) and 4(d)] show that the chemical composition of the defects is the same as that of the matrix region in the film and that the impurity oxygen concentration in the thin film is less than the detection limit of EDXS. We performed the EDXS line scans for six other vertical dislocations and nine domain boundaries, and obtained the same results. These results indicate that the defects are not an impurity phase such as $BaFeO_2$,[37] but may be edge or threading dislocations and/or domain boundaries. The defect sizes are estimated to be ~ 4 nm laterally and roughly double of the superconducting coherence length in the *ab* plane of $BaFe_2As_2$:P at 4 K,[7] which is consistent because such size defects effectively serve as vortex-pinning centers.

In summary, a high-self-field $J_c$ of 7 MA/cm$^2$ was obtained for $BaFe_2As_2$:P epitaxial films and a high $J_c$ of over 1 MA/cm$^2$ was maintained even with an applied magnetic field of 9 T. This $J_c$ value at 9 T is the highest value obtained so far for iron-based superconductor thin films. In addition, a highly isotropic high $J_c$ performance was obtained by decreasing the film growth rate, which introduced vertical dislocations along the *c*-axis that served as strong vortex-pinning centers.

**Acknowledgment**

This study was supported by the Japan Society for the Promotion of Science (JSPS), Japan, through the "Funding Program for World-Leading Innovative R&D on Science and Technology (FIRST Program)" and the Ministry of Education, Culture, Sports, Science and Technology (MEXT) Element Strategy Initiative to Form Core Research Center. H. Hiramatsu was also supported by a JSPS Grant-in-Aid for Young Scientists (A) Grant Number 25709058 and a JSPS Grant-in-Aid for Scientific Research on Innovative Areas "Nano Informatics" Grant Number 25106007.

**References**

1. Y. Kamihara, T. Watanabe, M. Hirano, and H. Hosono, J. Am. Chem. Soc. 130, 3296 (2008).

2. Z. -A. Ren, W. Lu, J. Yang, W. Yi, X. -L. Shen, Z. -C. Li, G. -C. Che, X. -L. Dong, L. -L. Sun, F. Zhou, and Z. -X. Zhao, Chin. Phys. Lett. 25, 2215 (2008).

3. F. Hunte, J. Jaroszynski, A. Gurevich, D.C. Larbalestier, R. Jin, A.S. Sefat, M.A. McGuire, B.C. Sales, D.K. Christen, and D. Mandrus, Nature 453, 903 (2008).

4. A. Yamamoto, J. Jaroszynski, C. Tarantini, L. Balicas, J. Jiang, A. Gurevich, D.C. Larbalestier, R. Jin, A.S. Sefat, M.A. McGuire, B.C. Sales, D.K. Christen, and D.

Mandrus, Appl. Phys. Lett. 94, 062511 (2009).

5. T. Katase, Y. Ishimaru, A. Tsukamoto, H. Hiramatsu, T. Kamiya, K. Tanabe, and H. Hosono, Nat. Commun. 2, 409 (2011).

6. H. Hiramatsu, T. Katase, T. Kamiya, and H. Hosono, J. Phys. Soc. Jpn. 81, 011011 (2012).

7. M. Miura, S. Adachi, T. Shimode, K. Wada, A. Takemori, N. Chikumoto, K. Nakao, and K. Tanabe, Appl. Phys. Express 6, 093101 (2013).

8. A. Sakagami, T. Kawaguchi, M. Tabuchi, T. Ujihara, Y. Takeda, and H. Ikuta, Physica C 494, 181 (2013).

9. S. Lee, J. Jiang, Y. Zhang, C.W. Bark, J.D. Weiss, C. Tarantini, C.T. Nelson, H.W. Jang, C.M. Folkman, S.H. Baek, A. Polyanskii, D. Abraimov, A. Yamamoto, J.W. Park, X.Q. Pan, E.E. Hellstrom, D.C. Larbalestier, and C.B. Eom, Nat. Mater. 9, 397 (2010).

10. T. Katase, H. Hiramatsu, T. Kamiya, and H. Hosono, Appl. Phys. Express 3, 063101 (2010).

11. B. Maiorov, T. Katase, I.O. Usov, M. Weigand, L. Civale, H. Hiramatsu, and H. Hosono, Phys. Rev. B 86, 094513 (2012).

12. C. J. van der Beek, M. Konczykowski, S. Kasahara, T. Terashima, R. Okazaki, T. Shibauchi, and Y. Matsuda, Phys. Rev. Lett. 105, 267002 (2010).

13. S. Demirdiş, Y. Fasano, S. Kasahara, T. Terashima, T. Shibauchi, Y. Matsuda, M. Konczykowski, H. Pastoriza, and C.J. van der Beek, Phys. Rev. B 87, 094506 (2013).

14. J.L. MacManus-Driscoll, S.R. Foltyn, Q.X. Jia, H. Wang, A. Serquis, L. Civale, B.

Maiorov, M.E. Hawley, M.P. Maley, and D.E. Peterson, Nat. Mater. 3, 439 (2004).

15. K. Matsumoto and P. Mele, Supercond. Sci. Technol. 23, 014001 (2010).

16. L. Fang, Y. Jia, C. Chaparro, G. Sheet, H. Claus, M.A. Kirk, A.E. Koshelev, U. Welp, G.W. Crabtree, W.K. Kwok, S. Zhu, H.F. Hu, J.M. Zuo, H.-H. Wen, and B. Shen, Appl. Phys. Lett. 101, 012601 (2012).

17. T. Katase, H. Hiramatsu, T. Kamiya, and H. Hosono, Supercond. Sci. Technol. 25, 084015 (2012).

18. See supplementary material at [http://dx.doi.org/10.1063/1.4875956] for the cross-sectional STEM image and EDXS spectra.

19. K. Iida, J. Hänisch, T. Thersleff, F. Kurth, M. Kidszun, S. Haindl, R. Hühne, L. Schultz, and B. Holzapfel, Phys. Rev. B 81, 100507 (2010).

20. S. Kasahara, T. Shibauchi, K. Hashimoto, K. Ikada, S. Tonegawa, R. Okazaki, H. Shishido, H. Ikeda, H. Takeya, K. Hirata, T. Terashima, and Y. Matsuda, Phys. Rev. B 81, 184519 (2010).

21. T. Katase, H. Hiramatsu, H. Yanagi, T. Kamiya, M. Hirano, and H. Hosono, Solid State Commun. 149, 2121 (2009).

22. H. Shishido, A.F. Bangura, A.I. Coldea, S. Tonegawa, K. Hashimoto, S. Kasahara, P.M.C. Rourke, H. Ikeda, T. Terashima, R. Settai, Y. Ōnuki, D. Vignolles, C. Proust, B. Vignolle, A. McCollam, Y. Matsuda, T. Shibauchi, and A. Carrington, Phys. Rev. Lett. 104, 057008 (2010).

23. K. Iida, J. Hänisch, R. Hühne, F. Kurth, M. Kidszun, S. Haindl, J. Werner, L.

Schultz, and B. Holzapfel, Appl. Phys. Lett. 95, 192501 (2009).

24. T. Kawaguchi, A. Sakagami, Y. Mori, M. Tabuchi, T. Ujihara, Y. Takeda, and H. Ikuta, Supercond. Sci. Technol. 27, 065005 (2014).

25. H. Hiramatsu, H. Sato, T. Katase, T. Kamiya and H. Hosono, to be published in Appl. Phys. Lett.

26. M. Miura, B. Maiorov, T. Kato, T. Shimode, K. Wada, S. Adachi, and K. Tanabe, Nat. Commun. 4, 2499 (2013).

27. K. Iida, J. Hänisch, C. Tarantini, F. Kurth, J. Jaroszynski, S. Ueda, M. Naito, A. Ichinose, I. Tsukada, E. Reich, V. Grinenko, L. Schultz, and B. Holzapfel, Sci. Rep. 3, 2139 (2013).

28. S. Lee, C. Tarantini, P. Gao, J. Jiang, J.D. Weiss, F. Kametani, C.M. Folkman, Y. Zhang, X.Q. Pan, E.E. Hellstrom, D.C. Larbalestier, and C.B. Eom, Nat. Mater. 12, 392 (2013).

29. K. Iida, S. Haindl, T. Thersleff, J. Hänisch, F. Kurth, M. Kidszun, R. Hühne, I. Mönch, L. Schultz, B. Holzapfel, and R. Heller, Appl. Phys. Lett. 97, 172507 (2010).

30. S. Trommler, J. Hänisch, V. Matias, R. Hühne, E. Reich, K. Iida, S. Haindl, L. Schultz, and B. Holzapfel, Supercond. Sci. Technol. 25, 084019 (2012).

31. V. Braccini, S. Kawale, E. Reich, E. Bellingeri, L. Pellegrino, A. Sala, M. Putti, K. Higashikawa, T. Kiss, B. Holzapfel, and C. Ferdeghini, Appl. Phys. Lett. 103, 172601 (2013).

32. H. Hiramatsu, T. Katase, Y. Ishimaru, A. Tsukamoto, T. Kamiya, K. Tanabe, and H.

Hosono, Mater. Sci. Eng. B 177, 515 (2012).

33. L.D. Cooley, P.J. Lee, and D.C. Larbalestier, Phys. Rev. B 53, 6638 (1996).

34. A. Godeke, Supercond. Sci. Technol. 19, R68 (2006).

35. C.G. Zhuang, S. Meng, H. Yang, Y. Jia, H.H. Wen, X.X. Xi, Q.R. Feng, and Z.Z. Gan, Supercond. Sci. Technol. 21, 082002 (2008).

36. H. Sato, T. Katase, W.N. Kang, H. Hiramatsu, T. Kamiya, and H. Hosono, Phys. Rev. B 87, 064504 (2013).

37. Y. Zhang, C.T. Nelson, S. Lee, J. Jiang, C.W. Bark, J.D. Weiss, C. Tarantini, C.M. Folkman, S.-H. Baek, E.E. Hellstrom, D.C. Larbalestier, C.-B. Eom, and X. Pan, Appl. Phys. Lett. 98, 042509 (2011).

**Figure captions**

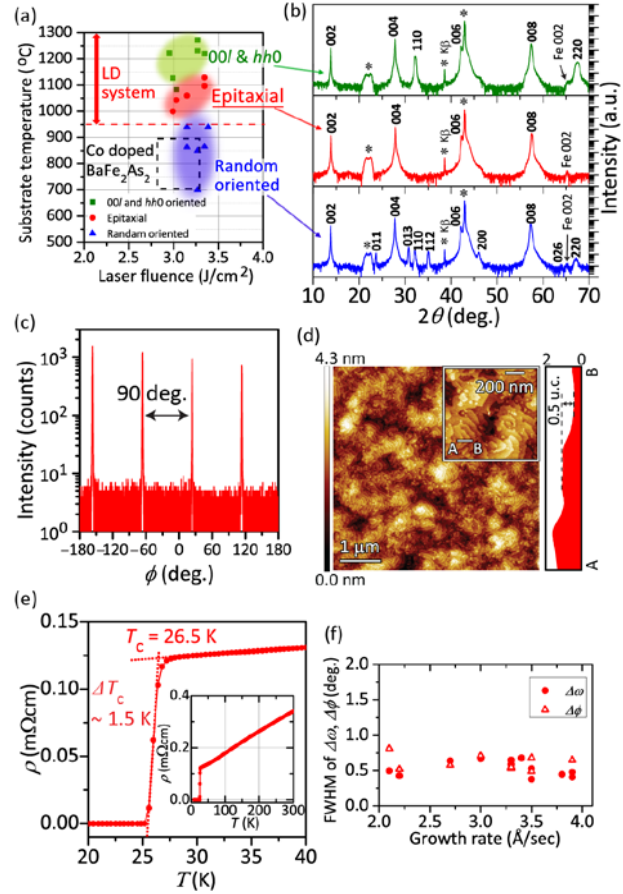

Fig. 1. (Color online) Growth of BaFe$_2$As$_2$:P films and structural and electrical properties of the epitaxial film. (a) Relationship between growth condition (substrate temperature $T_s$ and laser fluence) and crystallographic orientation of BaFe$_2$As$_2$:P films. Three regions with different orientation structures obtained with different $T_s$ values are shown. The dashed square region shows the optimum condition for epitaxial growth of BaFe$_2$As$_2$:Co.[17] (b) Typical XRD patterns, measured in Bragg–Brentano configuration, of BaFe$_2$As$_2$:P thin films for the three $T_s$ regions in panel (a). (c) In-plane $\phi$ scan of the 200 diffractions of the BaFe$_2$As$_2$:P thin film grown at the optimum $T_s$. (d) AFM scan showing surface morphology of BaFe$_2$As$_2$:P film grown at the optimum $T_s$. Upper-right inset shows a magnified image. A cross-sectional profile along the horizontal line A-B in the inset is shown to the right ("0.5 u.c." denotes the half size of the $c$-axis unit-cell length). (e) Resistivity as a function of temperature for BaFe$_2$As$_2$:P thin film grown at the optimum $T_s$. Inset shows data from 10 to 300 K. (f) Dependence of FWHM of $\Delta\omega$ and $\Delta\phi$ on growth rate of BaFe$_2$As$_2$:P films grown at the optimum $T_s$.

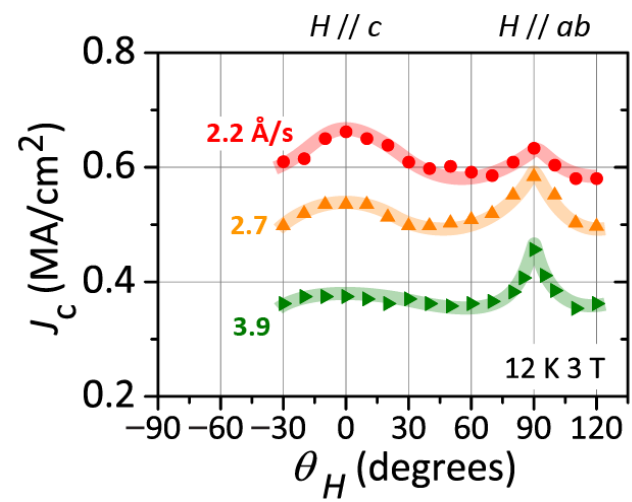

Fig. 2 (Color online) Critical current density $J_c$ as a function of angle of applied magnetic field $\theta_H$ for BaFe$_2$As$_2$:P epitaxial films grown at three growth rates, with $\mu_0 H$ = 3 T and $T$ = 12 K.

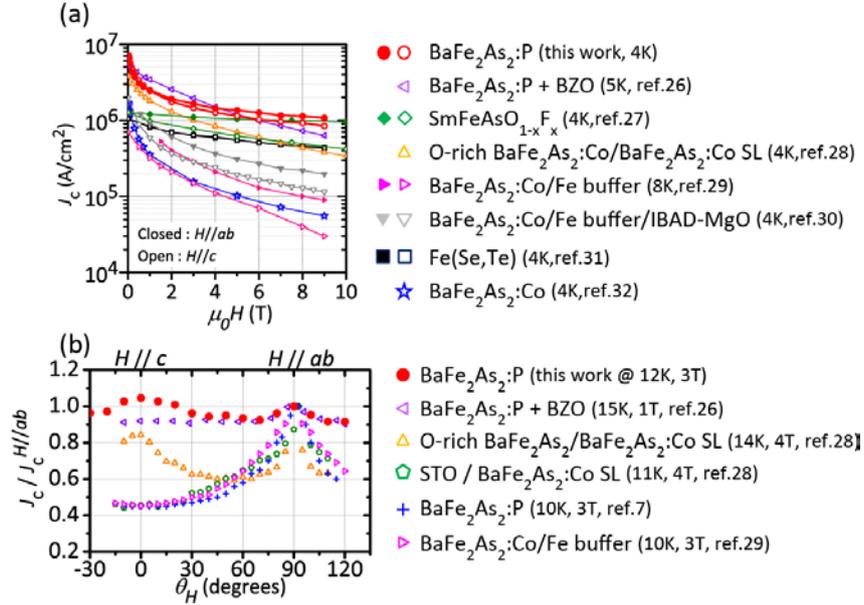

Fig. 3. (Color online) (a) Critical current density $J_c$ as a function of magnetic-field $H$ for BaFe$_2$As$_2$:P epitaxial film grown at the optimum $T_s$ (1050 °C) and growth rate (2.2 Å/s) (circles). Open and closed symbols show the data obtained by applying $H \parallel c$ and $H \parallel ab$, respectively. (b) Dependence of $J_c$ (normalized to $J_c^{H//ab}$) on angle of external magnetic field $\theta_H$ for BaFe$_2$As$_2$:P epitaxial film at 12 K and 3 T (circles). Also shown for comparison are the data for BaFe$_2$As$_2$:P + 3 mol% BaZrO$_3$ nanoparticles (left-pointing triangles),[26] SmFeASO$_{1-x}$F$_x$ (diamonds),[27] O-rich BaFe$_2$As$_2$/BaFe$_2$As$_2$:Co superlattice (SL) (triangles),[28] STO/BaFe$_2$As$_2$:Co SL (pentagons),[28] BaFe$_2$As$_2$:Co/Fe buffer (right-pointing triangles),[29] BaFe$_2$As$_2$:P (crosses),[7] BaFe$_2$As$_2$:Co/Fe/IBAD-MgO (inverse triangles),[30] Fe(Se,Te) (squares),[31] and BaFe$_2$As$_2$:Co (stars).[32]

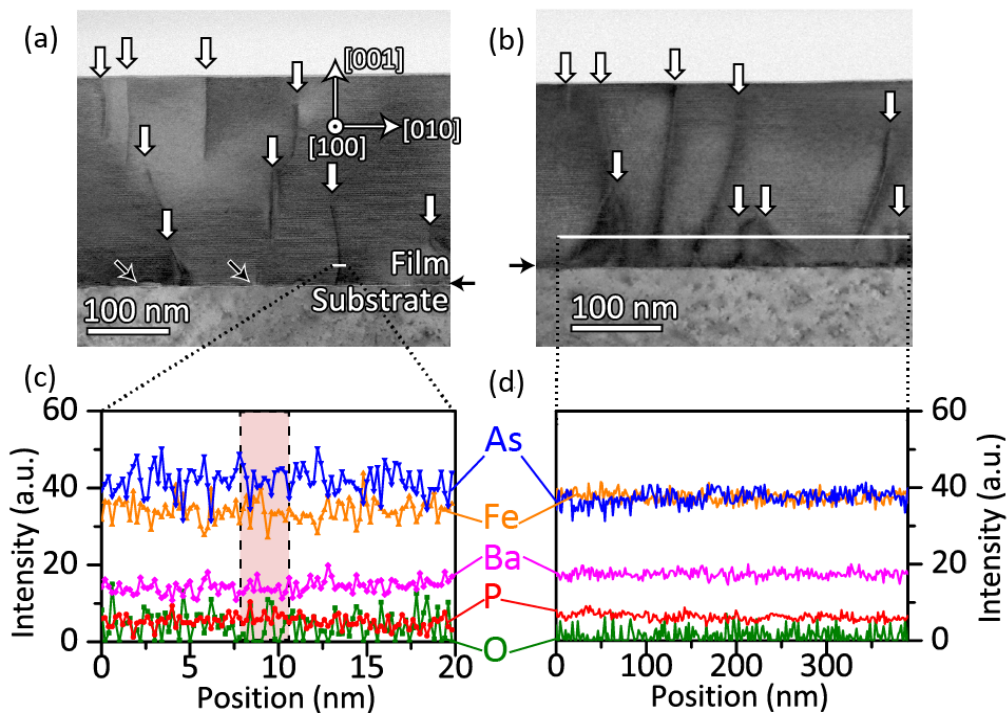

Fig. 4. (Color online) Cross-sectional bright-field STEM images of BaFe$_2$As$_2$:P epitaxial films grown at the growth rates (a) 2.2 Å/s and (b) 3.9 Å/s. The horizontal black arrow on the side of each figure indicates the heterointerface between the substrate and the BaFe$_2$As$_2$:P epitaxial film. The vertical white arrows indicate the vertical defects, and the slanted black arrows indicate oxygen-rich island structures. Panels (c) and (d) show the results of an STEM–EDXS line scan along the horizontal line shown in panels (a) and (b), respectively. The dashed shaded region in panel (c) corresponds to the vertical defect in panel (a).

# Supplementary material for " High critical-current density with less anisotropy in BaFe₂(As,P)₂ epitaxial thin films: effect of intentionally grown *c*-axis vortex-pinning centers"

Hikaru Sato[1], Hidenori Hiramatsu[1, 2], Toshio Kamiya[1, 2], and Hideo Hosono[1, 2, 3]

1: *Materials and Structures Laboratory, Tokyo Institute of Technology, Mailbox R3-1, 4259 Nagatsuta-cho, Midori-ku, Yokohama 226-8503, Japan*

2: *Materials Research Center for Element Strategy, Tokyo Institute of Technology, Mailbox S2-16, 4259 Nagatsuta-cho, Midori-ku, Yokohama 226-8503, Japan*

3: *Frontier Research Center, Tokyo Institute of Technology, Mailbox S2-13, 4259 Nagatsuta-cho, Midori-ku, Yokohama 226-8503, Japan*

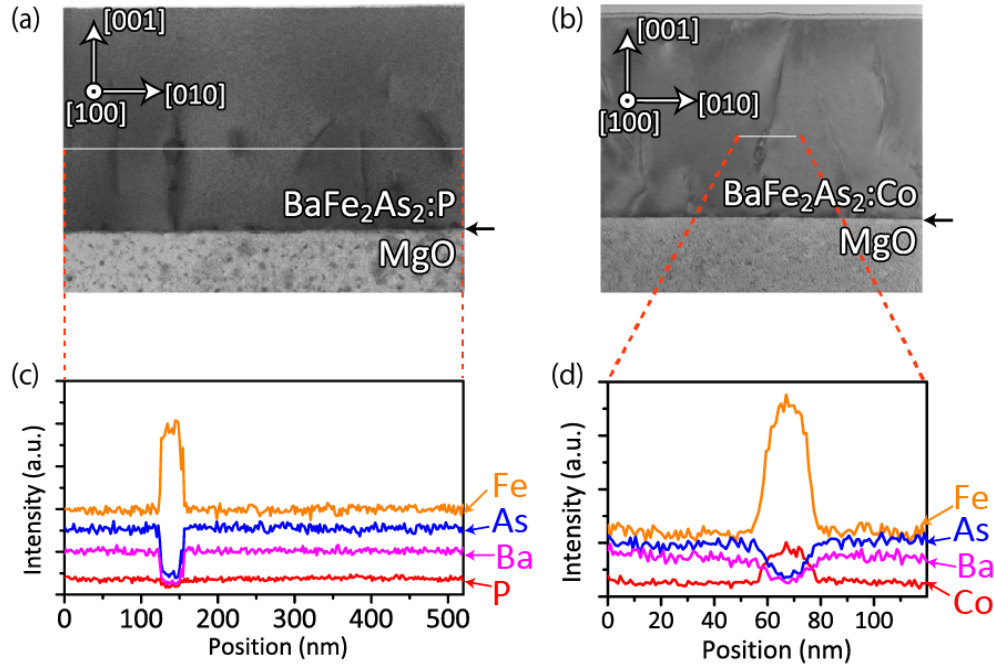

Supplementary FIG. S1. Cross-sectional bright-field STEM images of BaFe₂As₂:P (a) and BaFe₂As₂:Co (b)[Ref] epitaxial films. Horizontal arrows indicate the heterointerfaces between the films and MgO substrates. Panels (c) and (d) show the results of EDXS line-scans along the white horizontal lines in panels (a) and (b), respectively.

[Ref] T. Katase, H. Hiramatsu, T. Kamiya, and H. Hosono, Supercond. Sci. Technol. **25**, 084015 (2012).